# Integer defects in active nematics of ellipsoidal Quincke rollers

Bao Ying Fan, Lin Liu, Jie Wu, Chen Song,
Wen-de Tian, Kang Chen, and Tian Hui Zhang*

*Center for Soft Condensed Matter Physics and Interdisciplinary Research & School of Physical Science and Technology, Soochow University, Suzhou 215006, China*

Transverse propulsion changes topological ordering in dense ellipsoidal Quincke rollers. Under an alternating-current electric field, a polar velocity field coexists with perpendicular nematic particle alignment. The system supports integer +1 velocity vortices with radial particle axes and −1 saddle defects. Oppositely charged defects migrate along domain walls and annihilate as polar domains merge. These results identify propulsion direction relative to particle shape as a control parameter for defect textures and collective ordering.

*Introduction.*— Collective motion[1–4] is one of the most prominent manifestations of active matter, emerging in systems such as cytoskeletal assemblies,[5] bacterial suspensions,[6–8] animal groups,[9–11] and synthetic microswimmers.[12] A central theme in active-matter physics is to understand how microscopic propulsion and the interactions between active units determine macroscopic behaviors. To explore the general principles underlying various collective motions, observations have been conducted in many model systems. Among these, elongated bacteria and actin filaments,[13] modeled as collections of rod-like active particles,[14] have been widely studied both theoretically and experimentally as models of active nematics with long-range orientational order.[14] Active nematics represent a class of nonequilibrium materials with no equilibrium counterpart. More importantly, active nematics offer a robust model for understanding collective cell migration,[15] intracellular organization, and biological pattern formation.[16] The role of topological defects in active nematics is also of fundamental importance in controlling morphogenesis and extrusion events.[17,18]

In two-dimensional nematic phases,[14,19,20] the head-tail symmetry of the director field makes directions n and −n equivalent, and the lowest-energy topological defects are half-integer: comet-like $+1/2$ defects and trefoil-like $-1/2$ defects. Around $+1/2$ defects, the asymmetric distortion of the nematic field creates unbalanced active stresses that propel them as self-propelled particles. For the threefold-symmetric nematic field around $-1/2$ defects, the active stresses are balanced at their core, leading to zero net self-propulsion.[20–22] They typically move with surrounding flows. Defects with opposite charge annihilate or are created in pairs. The continuous creation, motion, and annihilation of defects leads to a chaotic, turbulent-like state even at low Reynolds numbers. The motion of $+1/2$ defects is a primary driver of the flow field, acting as local force dipoles that stir the fluid.[20]

So far, the popular understanding of topological defects and active nematics is based on the model in which active rod-like units are propelled along the long axis, such that the polar orientation of active rods and the directions of propulsion or velocity are intrinsically coupled.[14,15,23] In this case, active stress produces an active force that propels the $+1/2$ defect along the polar axis. Despite its fundamental importance, the question remains: if active rods are propelled perpendicular to their long axes, what kinds of topological defects will form and how will they move in active nematics? Here, ellipsoidal Quincke rollers are employed as a model system, in which active ellipsoids are driven transversely along their short axes.[24] Distinct from previous studies of active ellipsoidal Quincke rollers,[24–26] dense suspensions in this study develop large-scale polar order with integer topological defects rather than half-integer defects. These defect structures differ from those in the conventional polar phase, where +1 defects feature particles aligning tangentially. Instead, +1 defects in our system possess a radial vortex configuration, with particle long axes strictly pointing radially toward or away from the defect core, while −1 defects exhibit saddle-point structures with fourfold symmetry. Furthermore, domain walls provide pathways for topological defect motion. Driven by elastic attractive interactions between oppositely charged defects, +1 and −1 defects migrate along these domain walls toward each other and annihilate, substantially enhancing the efficiency of system ordering.

*Results.*— Ellipsoidal Quincke rollers driven by a direct-current (DC) electric field exhibit two different states. Near the critical field, the rollers exhibit strong chirality due to the tilted spinning axis and the resulting precession.[24] Increasing the spinning angular speed can suppress the precession. Distinct from rod-like particles propelled along the long axes, ellipsoidal Quincke rollers are propelled along the short axes, such that two rollers moving with opposite velocities block each other although their long axes are parallel, giving rise to clustering rather than aligning in collision.[24] Near the critical field, ellipsoidal Quincke rollers can align their velocities by flipping[25] to form a small-scale polar fluid with nematic order. However, flocking near the critical field is not stable because of the strong chirality. Recently, we found that a high-frequency alternating-current (AC) electric field can effectively suppress the chirality and facilitate formation of a large-scale polar fluid.[26]

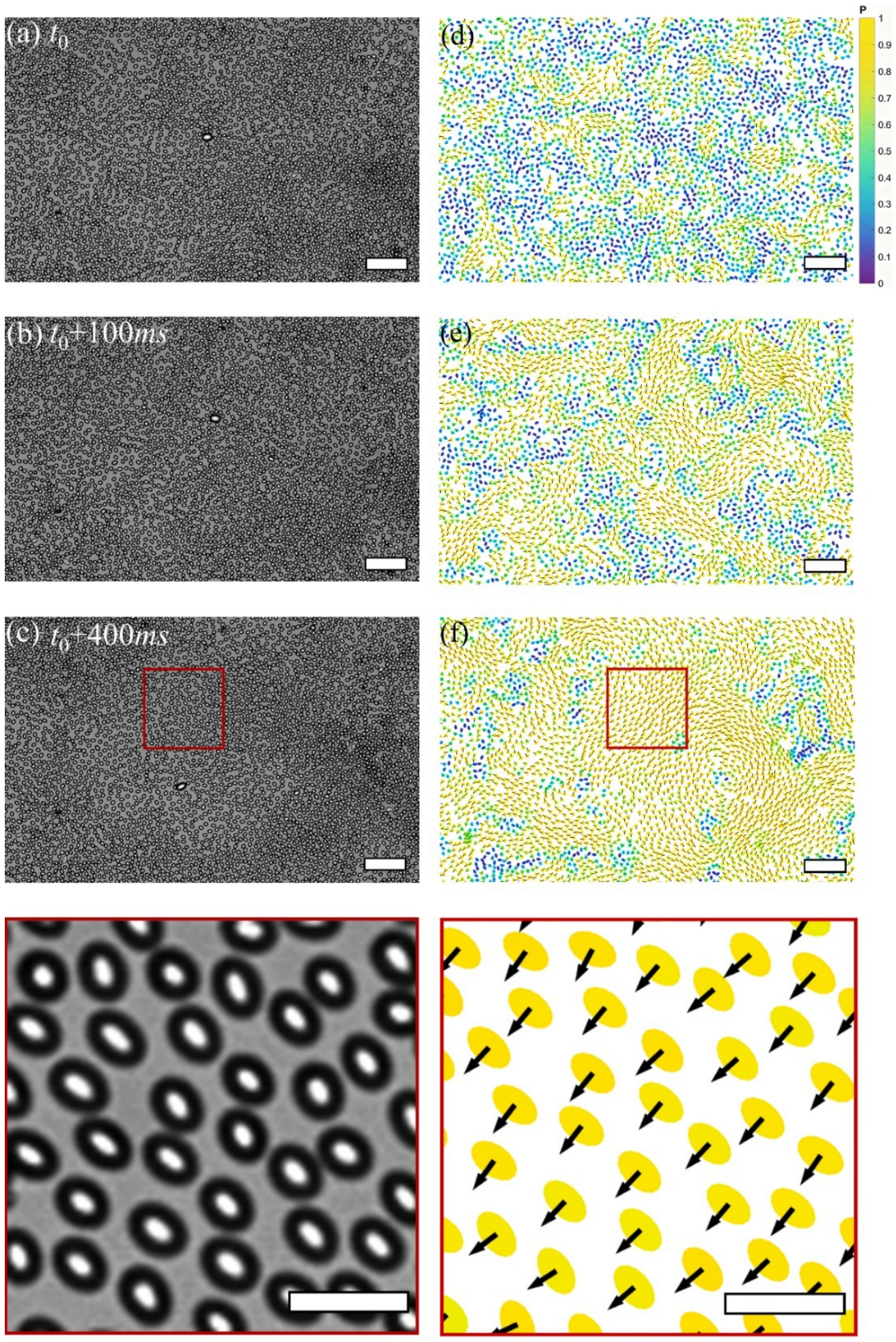

FIG. 1. The evolution process from a disordered state to a large-scale coherent state under an AC electric field. The field parameters were set as peak electric field Ep/Ec = 2.3, ground electric field Eg = 0, and frequency f = 130 Hz, where Ec is the critical electric field. (a–c) Snapshots of ellipsoidal rollers. (d–f) Evolution of the polar order parameter P corresponding to (a–c). Scatter colors represent the local polarization order parameter P, from blue (0, random) to yellow (1, highly aligned); black arrows denote velocity directions. Scale bars in (a–f), 100 μm; insets in (c) and (f), 20 μm.

Figure 1 exhibits a typical emergence of large-scale polar fluid of ellipsoidal Quincke rollers (Figs. 1(a)–1(c)). To quantify polar order, the local polar order parameter P is calculated from the velocities of individual rollers (see Materials and Methods). It shows that the initial polar order emerges locally as small-scale polar domains (Fig. 1(d)). The polar domains grow by aligning surrounding rollers (Fig. 1(e)). A globally ordered polar phase is achieved by merging polar domains (Fig. 1(f)). In the growth and merging of polar domains, flipping-mediated alignment of ellipsoidal rollers[26] plays a key role. Accompanying the formation of polar domains, ellipsoidal rollers are aligned along the long axis simultaneously. In this case, the nematic orientation (Fig. 1(c)) is perpendicular to the orientation of polar order (Fig. 1(f)).

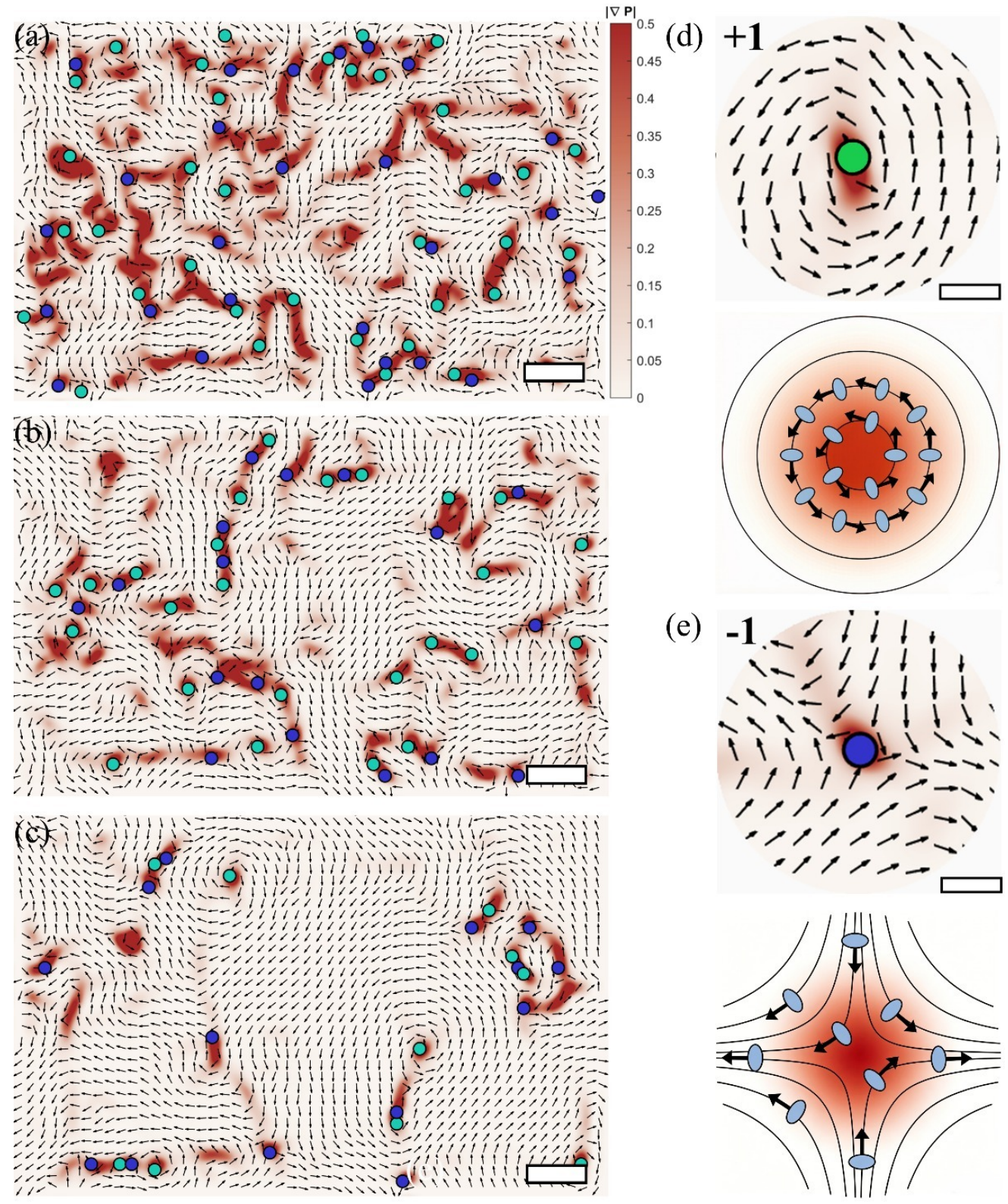

FIG. 2. Evolution of the local strain magnitude |∇P|. (a–c) Time evolution of the polarization strain field |∇P|. Brick-red regions indicate the maxima of |∇P|, corresponding to domain walls between differently oriented domains; black arrows denote local velocity vector directions. (d) +1 defect with a vortex-like core: the velocity vector rotates continuously by 2π around the center. (e) −1 defect with a saddle-point-like core and a hyperbolic flow manifold. Scale bars in (a–c), 100 μm; in (d–e), 50 μm.

At the boundaries between polar domains, the mismatch of polar orientations results in distortion of the polar field and therefore the formation of topological defects. To quantify spatial distortions of the polar field, the local strain magnitude |∇P| is defined and calculated from the velocities of individual rollers (see Materials and Methods). Ellipsoidal rollers in different domains are characterized by different directions of velocities as well as nematic orientations. Therefore, P experiences an abrupt change in the velocity direction across the boundaries between different domains, thereby producing extremely high polar-gradient values $|\nabla P|$. As a result, the gradient field $|\nabla P|$ offers a measurement of the magnitude of local distortion of the polar field.[27] The maximal points in the polar-gradient field locate the positions of topological defects. In the colored field of $|\nabla P|$, the centers of topological defects can be clearly identified at the domain walls. At the early state, as the number of defects is large, the interconnected domain walls form a network in the system (Fig. 2(a)). Accompanying the growth of the scale of polar order, the number of defects decreases due to annihilation and the network breaks up into segments (Figs. 2(b) and 2(c)). Two types of topological defects, rotational vortex-like (Fig. 2(d)) and diagonal saddle-like (Fig. 2(e)), are identified. In the vortex-like defects, the orientation of velocity or long axis of rollers changes $2\pi$ after a full-circle rotation (Fig. 2(d)), while in the saddle-like defects it changes $-2\pi$ (Fig. 2(e)). These two types of defects correspond to $+1$ and $-1$ defects. In a $+1$ defect, the velocity field forms a vortex while the long axes align radially. This structure contrasts sharply with the azimuthal alignment typically observed in cell systems.[28] The $-1$ defect exhibits a saddle-like flow structure with fourfold symmetry.

In the development of the globally ordered phase, the total defect number decreases monotonically (Fig. 3(a)) while the global polar order parameter increases (Fig. 3(b)). The synchronous decrease in the numbers of $+1$ and $-1$ defects indicates that defect reduction is not a random process, but rather a process of pairwise annihilation. Figure 4 shows that the motion of $\pm 1$ defects is confined to domain walls. The network of domain walls therefore restricts defect motion from two-dimensional planes to one-dimensional lines. Guided motion substantially enhances the efficiency of oppositely charged defects meeting and annihilating.

In active bacteria and cells, the $+1/2$ defect is propelled by unbalanced stress[29] while the $-1/2$ defect can move only due to interactions between defects. In our study, both $+1$ defects and $-1$ defects have symmetric structures, and the surrounding stresses are balanced. Therefore, they are not propelled and can move only through interactions. In the absence of propulsion, defect motion is confined within domain walls. Defects as singularities of the order-parameter field imply local orientational distortion, producing significant strain gradients and accumulating elastic energy in the system.[19] The system always tends to minimize the total elastic energy to reach a more stable state. When a pair of topologically opposite defects ($\pm 1$) coexist, their respective strain fields can partially cancel in space. By approaching each other, this pair of defects can significantly reduce the degree of elastic deformation in the entire system, thereby lowering the total free energy and giving rise to an attraction between $+1$ defects and $-1$ defects. Driven by attractive interaction, $+1$ defects and $-1$ defects tend to meet and annihilate, giving rise to larger polar domains (Fig. 4).

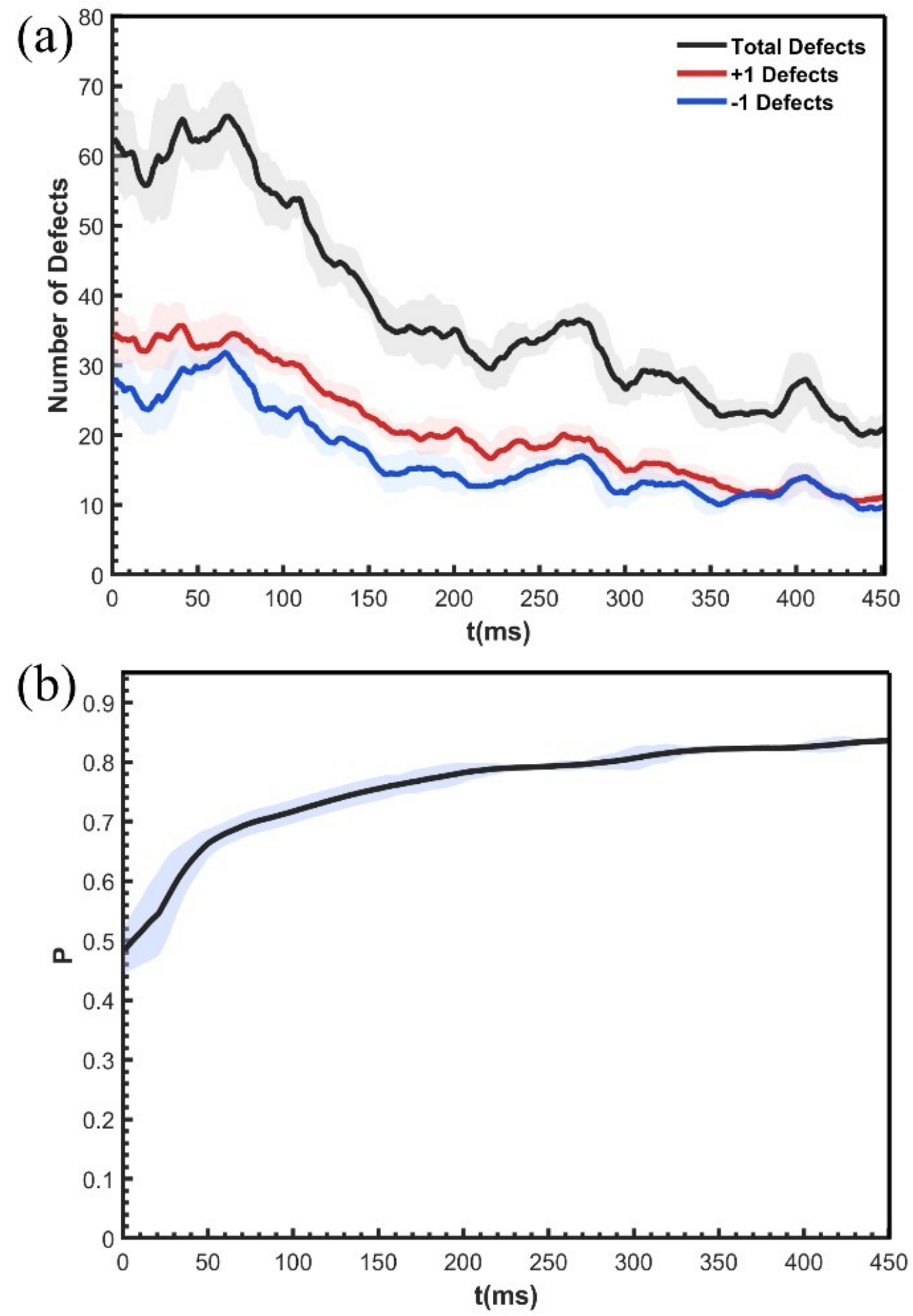

FIG. 3. Evolution curves of topological defect number and polarization order parameter over time. (a) Temporal evolution of the total number of topological defects (black), with numbers of $+1$ (red) and $-1$ (blue) defects; shaded regions are error bands. The close consistency of $+1$ and $-1$ defect numbers indicates pairwise annihilation. (b) Polarization order parameter increases over time; shaded region is the error band from 100 samples.

The above results show that the direction of propulsion relative to the particle's asymmetric shape is also a key control parameter for topological ordering in active nematics. Although the ellipsoidal rollers possess a nematic body axis, they propel perpendicular to this axis and form a polar velocity field. The system therefore combines transverse nematic alignment with polar motion, and its ordering dynamics is governed by integer $+1$ and $-1$ defects. This behavior differs from conventional active nematics, where propulsion and particle orientation are usually coupled and the dominant defects have charges $+1/2$ and $-1/2$.[14,29]

The polar $+1/2$ defect self-propels through asymmetric active stresses, promoting persistent defect creation, annihilation, and active turbulence. The $+1$ and $-1$ defects are not propelled due to the balanced stress, and their motion is confined along domain walls, where oppositely charged defects approach by attraction and annihilate.

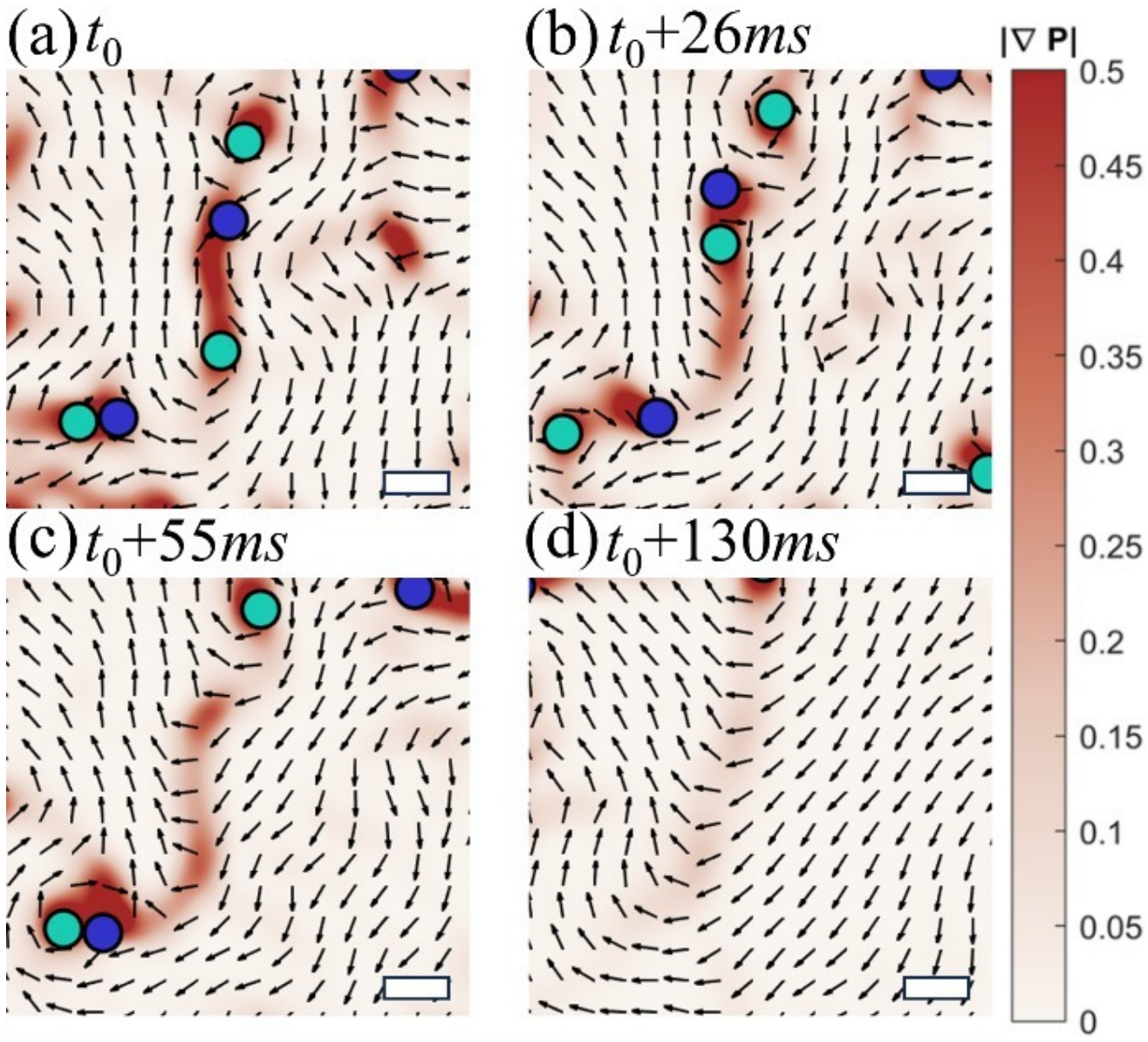

FIG. 4. Annihilation of ±1 topological defects on domain walls and reconfiguration of flow fields. (a) Multiple defect pairs are anchored on the distorted domain-wall network. (b–c) Oppositely charged defects approach along domain-wall trajectories and annihilate. (d) After annihilation, the originally disordered black arrows become aligned, and the corresponding domain walls weaken and vanish. Scale bars in (a–d), 50 μm.

The wall-guided annihilation resembles the topology-driven ordering reported for spherical Quincke rollers.[27] However, spherical rollers have no independent body axis. In our ellipsoidal system, the velocity and particle orientation remain perpendicular: a $+1$ velocity vortex is accompanied by radial alignment of the particle long axes, while the $-1$ defect exhibits a fourfold saddle texture. This microscopic structure is unique to transverse propulsion and cannot occur in isotropic roller flocks. Integer defects have also been studied in cell monolayers[28,30] and bacterial suspensions,[31] where they are commonly stabilized by confinement or patterned substrates and may split into half-integer defects. In contrast, the defects observed here arise spontaneously in the bulk, migrate along evolving domain walls, and disappear through pair annihilation.

In summary, dense ellipsoidal Quincke rollers driven perpendicular to their long axes develop large-scale polar order through defect-mediated coarsening. The polar velocity field coexists with a perpendicular nematic particle orientation and supports integer $+1$ vortex and $-1$ saddle defects. The $+1$ defects exhibit a distinctive structure in which the velocity circulates around the core while the particle long axes align radially. During ordering, oppositely charged defects migrate along domain walls and annihilate, weakening the walls and merging neighboring polar domains. The simultaneous decrease in defect number and increase in polar order directly link defect removal to global ordering. These results show that defect topology is determined not only by particle shape, but also by the direction of propulsion relative to that shape, providing a concrete design principle for anisotropic active colloids.

*Materials and Methods.*—

Preparation of ellipsoidal particles. The preparation follows the protocol established by Bricard et al.[32] Polystyrene spheres (Thermo Scientific G1000) with radius R = 10.1 μm are stretched into ellipsoidal particles according to Keville et al.[33] and Ho et al.[34]; despite variations in aspect ratio, all particles retain identical volume.

Experimental setup. Ellipsoidal Quincke particles with aspect ratio k ≈ 1.65 are dispersed in a 0.12 mol/L sodium bis(2-ethylhexyl)sulfosuccinate (AOT)/hexadecane solution and injected into a sealed cell assembled from two parallel ITO conductive glass plates. The effective cell thickness is 110 ± 10 μm, and the electric field is applied perpendicular to the electrode surface.

*Polar order parameter.* For each particle, P is computed from its velocity and those of its Voronoi neighbors:

$$P = \frac{1}{N}\left| \sum_{i=1}^{N} \frac{v_i}{|v_i|} \right| \tag{1}$$

*Local strain magnitude.* The local strain magnitude used to locate polar-field singularities is

$$|\nabla P| = \sqrt{\left(\frac{\partial p_x}{\partial x}\right)^2 + \left(\frac{\partial p_x}{\partial y}\right)^2} \tag{2}$$

*Supplemental Material.* Movie S1: development of large-scale active nematics. Ep = 2.3Ec, Eg = 0, and f = 130 Hz. The movie runs for 1929 ms of real time; the field of view is 800 × 520 $\mu m^2$. Scale bar: 100 μm.

*Acknowledgments.*— K.C., W.T., and T.H.Z. acknowledge financial support from the National Natural Science Foundation of China (Grant 12574230 to T.H.Z.; Grants 21774091 and 22673081 to K.C.; Grant 21674078 to W.T.).

*Data availability.*— The data supporting this article have been included as part of the Supplementary Information.

* zhangtianhui@suda.edu.cn